\documentclass[aps,twocolumn,notitlepage,superscriptaddress,floatfix]{revtex4}

\usepackage{amsmath,amssymb,amsthm,amsfonts}
\usepackage{graphicx,epsfig,color}
\usepackage{hyperref}
\usepackage[latin1,utf8]{inputenc}

\begin{document}

\title{BTZ gems inside regular Born-Infeld black holes}

\bigskip

\author{Christian G. B\"{o}hmer}
\email{c.boehmer@ucl.ac.uk}
\affiliation{Department of Mathematics, University College London, Gower Street, London, WC1E 6BT, United Kingdom.}

\author{Franco Fiorini}
\email{francof@cab.cnea.gov.ar} \affiliation{Departamento de Ingeniería en Telecomunicaciones (GDTyPE, GAIyANN) and Instituto Balseiro (UNCUYO), Centro Atómico Bariloche, Av. Ezequiel Bustillo 9500, CP8400, S. C. de Bariloche, Río Negro, Argentina.}



\begin{abstract}
  The regular black hole solution arising as a spherically symmetric vacuum solution of Born-Infeld gravity possesses an asymptotic interior structure which is very well described by a four dimensional generalization of the non-rotating BTZ metric. According to this picture no singularity exists, and instead, infalling observers experience a constant curvature manifold as they travel towards future null infinity. This is characterized by the BTZ event horizon. The exterior structure of the black hole is also studied, and it is shown that it corresponds to the Schwarzschild solution provided the black hole mass is not too small. In this way, the regular black hole state can be seen as a spacetime which connects two constant curvature asymptotic spaces, namely, the flat Minkowski spacetime in the outside region, and the locally AdS constant negative curvature one characterizing the BTZ-like asymptotic interior.
\end{abstract}

\maketitle

\section{Introduction}

A typical black hole solution in a gravitational theory is characterized by an event horizon that surrounds a geometrical singularity. Near this singularity our physical understanding breaks down and so does the mathematics involved. Singularities are actually very common within the context of metric theories of gravity and they are the rule rather than the exception, not only in General Relativity (GR)~\cite{Sin1,Sin2}, but in other theories of gravity as well. On the other hand, and by virtue of the increasing observational evidence concerning black hole physics~\cite{Bicep,Ghez,Abbot,Event}, singularities stopped being merely curiosities and started to take part in current central discussions as in the case of black hole entropy and the information loss paradox.

In the context of GR, one possible way to tackle the study of gravity in the strong curvature regime near black hole singularities, is either to postulate some sort of maximum curvature space replacing the innermost region of the black hole~\cite{Bardeen,Borde,Markov1,Markov2,Frolov,Brand1,Bronnikov:2001,Hayward:2005gi}, or by adding a matter content which makes gravity respond in a repulsive way in that regime~\cite{Eloy}. However, it should be remarked that replacing a singularity with a regular object is by no means straightforward and generally a trade-off has to be accepted. Often, the sources required for the construction of such solutions are difficult to motivate, or one has to consider matching spacetimes in a rather ad-hoc, non dynamical manner.

Motivated by the generally accepted fact that GR must be replaced by another theory when length scales become very small, we have recently addressed this topic by considering a GR modification with a somewhat different geometrical structure~\cite{Boehmer:2019uxv}. Our model is based on the teleparallel formulation of GR and its $f(T)$ generalization, which proved to be very successful in dealing with the Big Bang singularity featuring FLRW cosmological models~\cite{Ferraro:2006jd}. This generalization has the property that it is, in general, no longer invariant under arbitrary local Lorentz transformations in the strong field regime. Our specific Born-Infeld (BI) model introduces a new length scale $a_c$ which controls where the breakdown of the Lorentz invariance occurs. In particular the black hole singularity of `size' zero is replaced by a space of radius $a_c$ which is no longer singular. This constant curvature \emph{assembly space}, which is described by the interior Gott metric~\cite{Gott}, constitutes an asymptotic state lying at ingoing future null infinity. The replacement obtained is dynamically achieved, without the addition of any matter content nor topological change.

The main objective of the present work is to gain an in depth understanding of the structure of the spacetime \emph{near} the assembly space. We found, rather unexpectedly, that the line element near the assembly space is equivalent to a four dimensional extension of the non-rotating BTZ line element \cite{BTZ1,BTZ2} near its horizon. After a brief introduction into $f(T)$ gravity and the BI model in Section~\ref{glimpse}, we explicitly show in Section~\ref{interior} the coordinate transformations required to establish our result. The key idea is the use of null geodesics which leads to a type of Eddington-Finkelstein coordinates. To complete the geometrical picture, Section~\ref{exterior} shows that the solution near the black hole horizon is very well described by the Schwarzschild solution provided $r^2_{s} \gg a^2_{c}$, where $r_{s}=2GM$ is the Schwarzschild radius. We also show how the structure of the dynamical field -- the vierbein or \emph{tetrad} field $e^a$ -- results. The final conclusions are presented in Section~\ref{conclusions}. Throughout the paper, we will adopt the signature $-2$, and, as usual, Latin indexes $a:(0),(1),\ldots$ refer to tangent-space objects while Greek $\mu:0,1,\ldots$ denote spacetime components.

\bigskip

\section{A glimpse of $f(T)$-Born-Infeld gravity and its regular black hole}
\label{glimpse}

Before discussing the Born-Infeld gravitational scheme, it is convenient to briefly revisit $f(T)$ gravity, the general framework in which BI gravity is formulated. The action of $f(T)$ gravity can be considered a simple generalization of GR in its teleparallel (or absolute parallelism) formulation, and is written as
\begin{align}
  I=\frac{1}{16 \pi G }\int f(T)\,  e\, d^4x+I_{\rm matter}\,,
  \label{actionfT3D0}
\end{align}%
where $e=\sqrt{|\det g_{\mu\nu}|}$. There appears an arbitrary function $f$, at least twice differentiable, of the so called Weitzenb\"{o}ck scalar or torsion scalar $T = S^{a}{}_{\mu\nu} T_{a}{}^{\mu\nu}$ which is quadratic in torsion $T^{a}{}_{\mu\nu}=\partial_{\mu} e^{a}_{\nu}-\partial_{\nu} e^{a}_{\mu}$ by means of the 2-form $S^{a}{}_{\mu \nu}$ given by
\begin{align*}
  S^{a}{}_{\mu\nu} = \frac{1}{4}
  (T^{a}{}_{\mu\nu} - T_{\mu \nu }{}^{a} + T_{\nu \mu }{}^{a}) +
  \frac{1}{2} (\delta_{\mu}^{a} T_{\sigma\nu}{}^{\sigma} -
  \delta_{\nu}^{a} T_{\sigma\mu}{}^{\sigma})\,.
\end{align*}
The dynamical field underlying (\ref{actionfT3D0}) is the vierbein (tetrad field or coframe field) $e^{a}(x)=e^{a}_{\nu}(x)\,dx^\nu$, whose components in a local coordinate system are given by $e^{a}_{\nu}(x)$ and are functions of the coordinates. In the special case when $f$ is a linear function, we have the action
\begin{align}
  I=\frac{1}{16 \pi G }\int T\,  e\, d^4x+I_{\rm matter}\,,
  \label{actionGR}
\end{align}%
which is no other than the Hilbert-Einstein action in disguise; actually, it is relatively well known by now that the scalar curvature or Ricci scalar $R=g^{\mu\nu}R_{\mu\nu}$ (here $R_{\mu\nu}$ is the Ricci tensor), can be written as \cite{TEGRbook}
\begin{equation}
  \label{rels}
  R =-T + 2 e^{-1} \partial_{\nu }(e\,T_{\sigma }{}^{\sigma\nu}) \,,
\end{equation}
where $T_{\sigma }{}^{\sigma\nu}=g^{\mu\rho}T_{\mu\rho }{}^{\nu}$. Then, action (\ref{actionGR}) leads to Einstein's field equations written in a somewhat unconventional language, even though the equivalence is only valid provided the manifold has no boundary; in this case we can ignore the second term of the RHS of (\ref{rels}). It is worth mentioning the fact that the Weitzenb\"{o}ck scalar actually transforms as a scalar object only under general coordinate and global Lorentz transformations. Under a local Lorentz transformation of the tetrad $e^a=\Lambda^a_{b}(x)e^a$, torsion $T^a$ in the teleparallel setting does not transform covariantly; instead, the Weitzenb\"{o}ck scalar transforms modulo a surface term which does not affect the dynamics at the level of the GR field equations.

However, in the more general $f(T)$ action, the additional term emerging from $T$ when a Lorentz transformation acts on a given tetrad, is no longer a boundary term in the action. This means that $f(T)$ gravity is not locally Lorentz invariant in general, even though a local Lorentz transformation on shell subgroup \emph{remains} as a symmetry group. The additional degree of freedom  existing due to the breaking of the Lorentz symmetry \cite{RafaMajo1}, manifests itself as a particular arrangement of the tetrad field in the cotangent space $T^{\ast}\!\mathcal{M}$. This special arrangement conforms a parallelization of the manifold in consideration, which is actually defined by an equivalence class of tetrads connected by Lorentz transformations belonging to the \emph{remnant group} mentioned above. Details on this subtle subject can be consulted in \cite{Ferraro:2011,grupo,NosUn}.

The equations of motion or field equations are obtained from (\ref{actionfT3D0}) by varying with respect to the tetrad components $e^a_{\mu}$, and they are
\begin{eqnarray}
  \bigl(e^{-1}\partial_\mu(e\ S_a{}^{\mu\nu})+e_a^\lambda T^\rho{}_{\mu\lambda} S_\rho{}^{\mu\nu}\bigr)
  f^{\prime }+&\nonumber\\
  +\,S_a{}^{\mu\nu} \partial_\mu T f^{\prime \prime } - \frac{1}{4}
  e_a^\nu f =& -4\pi G e_a^\lambda \mathcal{T}_\lambda{}^{\nu},
  \label{ecuaciones}
\end{eqnarray}
where $\mathcal{T}_\lambda{}^{\nu}$ is the energy momentum tensor coming from the matter action $I_{\rm matter}$. Remarkably, though unsurprisingly because of the fact that $T$ contains just first derivatives of the tetrad field, equations (\ref{ecuaciones}) are of second order in derivatives of the vierbein, see also~\cite{Krssak:2018ywd}.

It could be argued that there is a considerable arbitrariness in the selection of the function $f(T)$ in (\ref{actionfT3D0}). Nonetheless, at least the same level of arbitrariness exists if one insists in relying on GR in situations where very strong curvatures arise, for instance `close' to strong curvature singularities such as the Big Bang or the ones present in the interior of several black holes solutions. Then, at least in these situations, there are good conceptual and theoretical reasons to consider other plausible $f(T)$ candidates. In particular, we will focus on
\begin{align}
  f(T)=\lambda[\sqrt{1+2T/\lambda}-1]\,,
  \label{BI}
\end{align}
where $\lambda$ is the BI parameter, whose units are of inverse length squared. This follow from the fact that the torsion scalar and the curvature scalar both have units of inverse length squared. Einstein's gravity is recovered in regions where $T/|\lambda| \ll 1$ which gives rise to the expansion $f(T) = T - T^2/\lambda + O(1/\lambda^2)$. The length scale $|\lambda|^{-1/2}$ is a measure of the scale at which local Lorentz invariance would no longer hold as a full symmetry. BI structures of the type~(\ref{BI}) were originally conceived in the context of electrodynamics for dealing with the singularity of the point-like electric charge at the origin \cite{BI} where the charge was located. The main idea was actually borrowed from special relativity; the relativistic lagrangian $\lambda_{0}[\sqrt{1+2L_{p}/\lambda_{0}}-1]$ with $\lambda_{0}=mc^2$, assures an upper bound for the particle speed $\dot{x}$ through the particle kinetic energy $L_{p}=m\dot{x}^2/2$. New physics will appear up when $\dot{x} \to c$ which is precisely the subject of study in special relativity.

Action~(\ref{BI}) was initially designed for treating the problem of the Big Bang singularity present in the standard FLRW cosmological model~\cite{Ferraro:2006jd}. After that, various extensions were studied in the same circumstances, even though they involve a different Lagrangian structure~\cite{yo2013,Vatu}. Within the (pseudo) Riemannian context, a number of BI proposals were studied as well~\cite{Deser,Fein2,Comelli,fder1,fder2,fder4,Max1,Max2,Cagri1}, see also Ref.~\cite{revOlmo} for a thorough account on the matter. However, many of these models are unable to deform vacuum GR solutions, and others lead to fourth-order field equations which are more difficult to study. These facts make them less relevant concerning the singularity resolution of vacuum GR black holes. Instead, the action coming from (\ref{BI}), i.e.
\begin{align*}
  I= \lambda/(16 \pi G) \int [\sqrt{1+2T/\lambda}-1]  e\, d^4x \,,
\end{align*}
has proved to contain a regular, spherically symmetric black hole interior solution, in pure vacuum when $\lambda<0$. In the following we summarize its main properties which were reported in~\cite{Boehmer:2019uxv}.

One of the key issues in $f(T)$ gravity is the obtention of a proper frame representing a given metric tensor. This is far from being obvious and can be considered the hardest step towards the characterization of a given solution. For instance, if we start from the \emph{closed} Kantowski-Sachs~\cite{KS66} line element
\begin{equation}
  \label{metks}
  ds^2=dt^2-b^2(t)dr^2-a^2(t)d\Omega^2 \,,
\end{equation}
a proper tetrad field is given by
\begin{align}
  e^{(t)} &= dt \,, \quad
  e^{(r)} = b(t)\cos\theta \,dr+a(t) \sin^2\theta \,d\phi \,,
  \label{parasixs2} \\
  e^{(\theta)} &= \sin\phi [b(t)\sin\theta dr+a(t) d\Omega_-]\,,
  \nonumber \\
  e^{(\phi)} &= \cos\phi [b(t)\sin\theta dr-a(t) d\Omega_+]\,.
  \nonumber
\end{align}
For ease of notation we introduced $d\Omega_+ = \tan\phi\, d\theta+\sin\theta\cos\theta\, d\phi$ and $d\Omega_- = \cot\phi\, d\theta-\sin\theta \cos\theta\, d\phi$. As we explained~\cite{Chileans1}, this is the proper frame for $f(T)$ gravity, adapted to the $\mathbb{R}^2 \times \mathbb{S}^2$ topology underlying (\ref{metks}). Other frames are suitable as well, and they are related to (\ref{parasixs2}) by means of Lorentz transformations belonging to the remnant group. Of course, if $f(T)=T$ every frame leading to (\ref{metks}) is equally valid, because TEGR is a theory for the metric tensor alone, and the local orientation of the tetrad becomes irrelevant. The Schwarzschild interior solution in KS form involves the scale factors $a(t)=a_{\rm KS}(t)$ and $b(t)=b_{\rm KS}(t)$ which are (implicitly) written as
\begin{equation}
  b_{\rm KS}(t) =b_{0}\,\tan(\eta(t)) \, \quad
  a_{\rm KS}(t) = 2M\,\cos\negmedspace^2(\eta(t)) \,.
\label{func-a}
\end{equation}
The time coordinate $t$ is parameterized by $\eta$ according to
\begin{equation}
  \label{tiempo_raro}
  t-t_{0} = 2M (\eta+\sin\eta\cos\eta) \,.
\end{equation}

The regular black hole interior is a high energy deformation of the Schwarzschild solution in the KS form. If we evaluate the vacuum $f(T)$ equations (\ref{ecuaciones}) in the frame (\ref{parasixs2}), they take the form
\begin{align}
  f+4f'(H_{a}^2+2H_{a}H_{b}) &= 0 \,,
  \label{Friedmann} \\
  f''H_{a}\dot{T}+f'(H_{a}H_{b}+2H_{a}^{2}+\dot{H}_{a})+f/4 &=0 \,.
  \label{sec-esp-1}
\end{align}
The Weitzenb\"{o}ck scalar associated to the frame (\ref{parasixs2}) is $T=-2(-a^{-2}+H_{a}^2)-4H_{a}H_{b}$. For the KS case, using (\ref{func-a}) one obtains $T_{\rm KS}=4a_{KS}^{-2}=M^{-2} \cos^{-4}(\eta(t))$. This clearly diverges as $\eta(t)\rightarrow \pi/2$, which in view of (\ref{func-a}), represents the Schwarzschild singularity. In turn, we have shown in \cite{Boehmer:2019uxv} that for the the BI theory $T$ results in
\begin{align}
  \label{scalartbi}
  T_{BI}=\frac{4}{a^2(t)}\Bigl(1-\frac{2}{|\lambda|a^2(t)}\Bigr) \,,
\end{align}
where the function $a(t)$ is a solution of the field equations~(\ref{Friedmann}) and~(\ref{sec-esp-1}). $T_{BI}$ now reaches a maximum $T_{\rm max}=|\lambda|/2$ when $a=a_{c}=2/\sqrt{|\lambda|}$, which defines what we called, the previously mentioned, \emph{assembly space}. We showed that the metric corresponding to this assembly space is exactly the one corresponding to the interior of an infinitely long (in the $Z$-direction) cosmic string
\begin{align}
  \label{metsobreacritfin}
  ds^{2}=dT^{2}-dZ^{2}- a_{c}^{2} d\Omega^{2}.
\end{align}
This metric was first obtained in \cite{Gott}, and it has constant curvature invariants
\begin{equation}
  \label{escalares}
  R = -2/a_{c}^{2},\quad
  R^{(2)} = 2/a_{c}^{4},\quad
  K= 4/a_{c}^{4},
\end{equation}
where $R^{(2)} = R_{\mu\nu}R^{\mu\nu}$, $K = R^{\alpha}{}_{\mu\nu\rho}R_{\alpha}{}^{\mu\nu\rho}$. We found that the assembly space (\ref{metsobreacritfin}) cannot be reached in a finite proper time. It represents an asymptotic inner spacetime where the radius $a(t)$ of the two-spheres in (\ref{metks}) goes to $a_{c}$ as $t\to \infty$. Then, the Schwarzschild singularity is replaced by an asymptotic spacetime of constant curvature. This regularization was obtained without invoking any matter content nor topological change.

As we have shown in~\cite{Boehmer:2019uxv}, we can combine the field equations in such a way that the Hubble function $H_b$ is eliminated. Hence one arrives at a single, nonlinear ODE for the scale factor $a(t)$ in the BI case. This equation reads
\begin{align}
  \Big[1-\frac{4}{|\lambda| a^{2}}\Big]
  \Big[\frac{1}{a^{2}}\Big(1-\frac{4}{|\lambda| a^{2}}\Big)+3 H_{a}^{2}+2\dot{H}_{a}\Big] -
  \frac{16 H^{2}_{a}}{|\lambda| a^{2}}=0 \,.
  \label{ecparares}
\end{align}
One of the key results of~\cite{Boehmer:2019uxv} was to find an approximate solution to the BI motion equations \emph{near} the assembly space. It ended up being
\begin{align}
  ds^2 &= dt^2 - b_0^2 t^2 dz^2 - a_c^2[1+2A \exp(-t^2/2a_c^2)]d\Omega^2 \,,
  \label{n1}
\end{align}
where $b_0$ and $A \ll 1$ are two non-vanishing integration constants. It is worth mentioning that expression (\ref{n1}) is valid for all times $t$ because the Gaussian function is bounded. It is the main purpose of this work to study (\ref{n1}) in detail.

\section{Asymptotic interior structure of the black hole: a BTZ-like description}
\label{interior}

In the following we will show that the regular black hole near the assembly space is described by the BTZ metric \footnote{The BTZ solution is actually obtained from (\ref{n2}) by taking constant $\phi$ hypersurfaces} in rather unusual coordinates, i.e., that metric (\ref{n1}) can be written as
\begin{align}
  ds_{\rm BTZ}^2 = (-\tilde{M}+\ell^{-2}R^2)dT^2 - \frac{dR^2}{(-\tilde{M}+\ell^{-2}R^2)} -
  R^2 d\Omega^2 \,,
  \label{n2}
\end{align}
provided $A \ll 1$. Here $\tilde{M}$ is the mass parameter and $\Lambda = -\ell^{-2}$ is the (negative) cosmological constant. To uncover this, we will introduce coordinates which are constant along null geodesics, similar to the familiar Eddington-Finkelstein coordinates. Let us consider radial null geodesics which means setting $d\Omega=0$ in (\ref{n1}), hence radial null geodesics satisfy $0 = dt^2 - b_0^2 t^2 dz^2 = t^2 (dt^2/t^2 - b_0^2 dz^2) = t^2 (dt/t - b_0 dz)(dt/t + b_0 dz)$. Consequently, we are led to introduce the coordinates
\begin{align}
  dU = dt/t - b_0 dz \quad \Rightarrow \quad
  U = \log t - b_0 z \,,\nonumber \\
  dV = dt/t + b_0 dz \quad \Rightarrow \quad
  V = \log t + b_0 z \,.
  \label{n4}
\end{align}
This allows us to write the $(t,z)$-part of the line element~(\ref{n1}) in the form
\begin{align*}
  \exp(U+V)dU dV = t^2 (dt^2/t^2 - b_0^2 dz^2) = dt^2 - b_0^2 t^2 dz^2 \,.
\end{align*}
Now we can write $\exp(U+V)=\exp(U)\exp(V)$ and introduce $d\tilde{U}=\exp(U)dU$ which means $\tilde{U} = \exp(U)$ and likewise for $\tilde{V}$. Then $t^2=\tilde{U}\tilde{V}$ and
\begin{align*}
  ds^2 &= dt^2 - b_0^2 t^2 dz^2 - a_c^2[1+2A \exp(-t^2/2a_c^2)]d\Omega^2
  \\ &=
  d\tilde{U} d\tilde{V}- a_c^2[1+2A \exp(-\tilde{U}\tilde{V}/2a_c^2)]d\Omega^2
\end{align*}
Finally we rescale $u = \tilde{U}/a_c$ and $v = \tilde{V}/a_c$ so that
\begin{align*}
  ds^2 = a_c^2\Bigl[du dv - \Bigl(1+2A \exp(-uv/2)\Bigr)d\Omega^2 \Bigr] \,.
\end{align*}
Since the $t=0$ coordinate singularity occurring in (\ref{n4}) corresponds to small values of the product $uv$, we can expand the exponential in the usual way $\exp(-uv/2) = 1 - uv/2 + \ldots$. Therefore, by neglecting higher order terms, we can write the interior solution near the assembly space (\ref{n1}) as
\begin{align}
  ds^2 = a_c^2\Bigl[du dv - \Bigl(1+2A(1-uv/2)\Bigr)d\Omega^2 \Bigr] \,.
  \label{n7a}
\end{align}
Let us emphasize that we are still free to re-scale either of the coordinates $\{u,v\}$ or $\{\tilde{u},\tilde{v}\}$; this will be important in what follows in order to identify the constants $a_c$ and $A$ with $\ell$ and $M$, respectively. To do so we re-scale $\{u,v\} \to \{\sigma u, \sigma v\}$ in (\ref{n7a}). Hence we have the final form of the metric near the assembly space
\begin{align}
  ds^2 &= a_c^2\Bigl[\sigma^2 du dv - \Bigl( 1+2A - A \sigma^2 uv\Bigr)d\Omega^2 \Bigr] \,.
  \label{fin1}
 \end{align}
On the other hand, we look at radial null geodesics in the BTZ line element (\ref{n2}); we introduce the coordinates
\begin{align}
  \bar{U} = \frac{\ell}{\sqrt{\tilde{M}}} \mathrm{arctanh}\bigl(R/\ell\sqrt{\tilde{M}}\bigr) + T \,, \nonumber\\
  \bar{V} = \frac{\ell}{\sqrt{\tilde{M}}} \mathrm{arctanh}\bigl(R/\ell\sqrt{\tilde{M}}\bigr) - T \,,
  \label{n8}
\end{align}
so that we have
\begin{align*}
  d\bar{U} d\bar{V} &= - dT^2 + \frac{dR^2}{(-\tilde{M}+\ell^{-2} R^2)^2} \,,
  \\
  (-\tilde{M}+\ell^{-2}R^2) &= -\frac{\tilde{M}}{\cosh^2(\sqrt{\tilde{M}}(\bar{U}+\bar{V})/2/\ell)} \,.
\end{align*}
In these coordinates the BTZ line element (\ref{n2}) becomes
\begin{multline}
  ds_{\rm BTZ}^2 =
  \frac{\tilde{M}}{\cosh^2\Bigl(\frac{\sqrt{\tilde{M}}(\bar{U}+\bar{V})}{2\ell}\Bigr)}
  \Bigl[ d\bar{U} d\bar{V} \\-
  \ell^2 \Bigl\{\cosh^2\Bigl(\frac{\sqrt{\tilde{M}}(\bar{U}+\bar{V})}{2\ell}\Bigr)-1\Bigr\} d\Omega^2\Bigr]
  \label{n11}
\end{multline}
Now, similar to before, we set $\sqrt{\tilde{M}}\bar{U}/\ell = \bar{u}$ and $\sqrt{\tilde{M}}\bar{V}/\ell$, this yields the result
\begin{align*}
  ds^2_{\rm BTZ} &=
  \ell^2 \Bigl[
    \frac{d\bar{u} d\bar{v}}{\cosh^2\bigl(\frac{\bar{u}+\bar{v}}{2}\bigr)} -
    \tilde{M} \Bigl(1 - \frac{1}{\cosh^2\bigl(\frac{\bar{u}+\bar{v}}{2}\bigr)}\Bigr) d\Omega^2
    \Bigr] \,,
\end{align*}
where the bars were introduced in order to distinguish the coordinates from the previous case. The BTZ event horizon is located at $R=\ell\sqrt{\tilde{M}}$, which corresponds to the argument of the function $\mathrm{arctanh}$ becoming $1$ in (\ref{n8}). Therefore we identify $R \rightarrow \ell\sqrt{\tilde{M}}$ with $\bar{U} \rightarrow \infty$ and $\bar{V} \rightarrow \infty$ which also holds for $\bar{u},\bar{v}$. Hence we can make the standard approximation $\cosh^{-2}((\bar{u}+\bar{v})/2) \approx 4\exp(-(\bar{u}+\bar{v})/2) = 4\exp(-\bar{u}/2)\exp(-\bar{v}/2)$ for large arguments of the function. In turn, this allows us to write the BTZ solution as
\begin{multline}
  ds^2_{\rm BTZ} = \ell^2 \Bigl[
    4 \exp(-\bar{u}/2)\exp(-\bar{v}/2) d\bar{u} d\bar{v} \\-
    \tilde{M} \Bigl(1 - 4 \exp(-\bar{u}/2)\exp(-\bar{v}/2) \Bigr) d\Omega^2
    \Bigr]\,.
  \label{eqn:BTZ2}
\end{multline}
Finally we introduce $\tilde{u}=-4\exp(-\bar{u}/2)$ and $\tilde{v}=-4\exp(-\bar{v}/2)$ to arrive at
\begin{align}
  ds^2_{\rm BTZ} &= \ell^2 \Bigl[
    d\tilde{u}d\tilde{v} -
    \tilde{M} \Bigl(1 - \tilde{u}\tilde{v}/4 \Bigr) d\Omega^2
    \Bigr] \,.
  \label{eqn:BTZ3}
\end{align}
We now can finally compare the expressions (\ref{fin1}) and (\ref{eqn:BTZ3}). The following conditions are implied for identifying the two line elements:
\begin{align*}
  a_c^2 \sigma^2 = \ell^2\,, \quad a_c^2 (1+2A) = \ell^2 \tilde{M}\,, \quad
  a_c^2 A \sigma^2 = \ell^2 \tilde{M}/4 \,.
\end{align*}
Clearly we have $a_c \sigma = \ell$ so that the remaining two conditions become
\begin{align*}
  (1+2A) = \sigma^2 \tilde{M}\,, \quad
  A = \tilde{M}/4 \,,
\end{align*}
then, $\sigma^2 = (1+2A)/4A$. This completes the identification as we can finally state
\begin{align}
  \label{idenfin}
  \ell = a_c \sqrt{(1+2A)/(4A)}\,, \quad \tilde{M} = 4 A \,.
\end{align}

Taking into account the previous relations we conclude
\begin{align*}
  R_h = \ell\sqrt{\tilde{M}} = a_c \sqrt{1+\tilde{M}/2} \approx a_c(1+\tilde{M}/4) \,,
\end{align*}
for small values of $A=\tilde{M}/4$, which we assumed throughout. This means that the critical radius defining the assembly space, approximately corresponds to the horizon of the BTZ solution for small masses. Actually, the BTZ horizon coincides with $a_{c}$ only when $\tilde{M}=0$. So, naturally the question arises: are there two surfaces of interest? Clearly $R_h>a_c$ for a positive $A=\tilde{M}/4$, so that one may wrongly conclude that the horizon at $R_h$ surrounds the critical surface. This is not the case, because the assembly space is not a solution of the field equations in itself, it is not part of the spacetime at all; this can be easily checked by inserting $a=a_{c}$, $H_{a}=0$ in (\ref{Friedmann}) and (\ref{sec-esp-1}). In fact, they read $f(T_{\rm max})=f(|\lambda|/2)=0$ (we used (\ref{scalartbi})), which is not consistent with
\begin{align*}
  f(T_{\rm max})=-|\lambda|\Big[\sqrt{1-2T_{\rm max}/|\lambda|}-1\Big]=|\lambda| \,,
\end{align*}
because $\lambda$ is different from zero.

The constant $A$ in line element (\ref{n1}) arose as a constant of integration when solving the equation (\ref{ecparares}) to first order using the expansion $a(t)=a_c(1+ \varepsilon F_1(t))$. However, when this procedure is extended to second order using the expansion $a(t)=a_c(1+ \varepsilon F_1(t) + \varepsilon^2 F_2(t))$ one can also solve explicitly the second order equations which contain additional constants of integration. It is possible to choose these constants such that $a(t=0)=a_c$ so that the \emph{location} of the critical surface is unaffected by the use of an expansion near $a_{c}$. It is worth mentioning that to expand near $a_{c}$ is well justified because the assembly space is a constant curvature space, see (\ref{escalares}).

The BTZ-like metric leads to non constant curvature invariants, which tend to the ones corresponding to the assembly space when evaluated at the BTZ horizon in the small mass limit. For instance, by evaluating the scalar curvature $R_{\rm BTZ}$ of (\ref{n2}) at the horizon, using the identifications (\ref{idenfin}), we get
\begin{align*}
  R_{\rm BTZ}\Bigr|_{h} =-\frac{2}{a_c\,\sqrt{1+\tilde{M}/2}}
  \Big(1-\frac{10 \tilde{M}}{a_c\,\sqrt{1+\tilde{M}/2}}\Big) \,,
\end{align*}
which tends to $R =-2/a_c$ when the mass $\tilde{M}$ goes to zero, in agreement with the value corresponding to the assembly space, see. Eq. (\ref{escalares}). In this way, we can interpret the assembly space as being the metric at future null infinity, the asymptotic state of any infalling observer entering the black hole. This is a most remarkable result, because the fearsome Schwarzschild singularity is no longer a threat.

\section{Near horizon and asymptotic exterior structure}
\label{exterior}

Unlike the high curvature regime characterizing the environs of the assembly space, the black hole near the horizon behaves as a Schwarzschild black hole to a high degree of accuracy. This can be shown easily by analyzing the contribution of the $\lambda$-terms in (\ref{ecparares}), near the Schwarzschild horizon. Bearing this in mind, let us introduce the Hubble function and its time derivative associated to $a_{\rm KS}(t)$. They are
\begin{align}
  \label{haches}
  H_{a\,{\rm KS}} &= -\frac{\tan(\eta)}{M(1+\cos(2\eta))}\,,\nonumber \\
  \dot{H}_{a\,{\rm KS}} &=\frac{\sec^6(\eta)}{2M^2}(-2+\cos(2\eta)) \,,
\end{align}
where we used $\dot{\eta}=1/(2M(1+\cos(2\eta)))$ which follows from the implicit definition of $\eta(t)$ given in (\ref{tiempo_raro}). Inspection of the terms appearing in~(\ref{ecparares}), $a_{\rm KS}$, $H_{a\, {\rm KS}}$ and $\dot{H}_{a\,{\rm KS}}$ as given by Eqs.~(\ref{func-a}) and (\ref{haches}), the terms involving $\lambda$ are proportional to
\begin{align*}
  \frac{1}{|\lambda| a_{\rm KS}^{2}} &\propto \frac{\sec^4(\eta)}{M^2|\lambda|}\,, \\
  \frac{H^{2}_{a\,{\rm KS}}}{|\lambda| a_{\rm KS}^{2}} &\propto
  \frac{\sec^8(\eta)\tan^2(\eta)}{M^4|\lambda|}\,, \\
  \frac{\dot{H}_{a\,{\rm KS}}}{|\lambda| a_{\rm KS}^{2}} &\propto
  \frac{\sec^{10}(\eta)(\cos(2\eta)-2)}{M^4|\lambda|} \,.
\end{align*}
As $\eta \to 0$, which represents the Schwarzschild horizon, the terms behave as $1/M^2|\lambda|$ and $1/M^4|\lambda|$. They are negligible as long as  $M^2 \gg  1/|\lambda|$, or $r^2_{s} \gg a^2_{c}$, which means that the spacetime in the vicinity of the horizon is very well described by the Schwarzschild solution as long as the black hole has not lost too much mass due to evaporation. On the contrary, late evaporation stages involve smaller masses and $\eta\rightarrow\pi/2$, where $\tan(\eta)$ and $\sec(\eta)$ diverge. This is the regime in which the BI black hole strongly departs from the Schwarzschild one.

The standard KS coordinate change will bring the metric (\ref{metks}) into the familiar Schwarzschild interior form. Let us treat $a_{\rm KS}(t)$ as a new coordinate and write $a_{\rm KS}(t)=a$ from now on. Then $\eta(t)=\eta(a)$ and $dt/da = (\partial t/\partial \eta)/(\partial\eta/\partial a)$. According to (\ref{tiempo_raro}), we have
\begin{align*}
  \frac{\partial t}{\partial \eta}=2M(1+\cos(2\eta))=4M\cos^2(\eta)=2a \,,
\end{align*}
in view of (\ref{func-a}). Inverting Eq.~(\ref{func-a}) we have
\begin{align*}
  \eta(a)=\arccos(\sqrt{a/2M}) \,, \quad
  \frac{\partial\eta}{\partial a}=\frac{-1}{2a\sqrt{-1+2M/a}} \,,
\end{align*}
from which is clear that the coordinate change is valid if $a<2M$. Finally we get
\begin{align}
  \label{kschange4}
  \frac{dt}{da}=\frac{-1}{\sqrt{-1+2M/a}}\,,\quad
  dt^2=\frac{-\,da^2}{(1-2M/a)} \,.
\end{align}
It only remains to write $b(t)$ as a function of the new coordinate $a$. Using (\ref{func-a}) we have
\begin{align*}
  b(t)=b_{0}\tan\Big(\arccos(\sqrt{a/2M})\Big)=\frac{\sqrt{1-a/2M}}{\sqrt{a/2M}} \,,
\end{align*}
which then reads $b^2(t)=2M/a-1$, where we have eliminated $b_{0}$ with a trivial coordinate transformation. We arrive at the KS metric (\ref{metks}) in the familiar form
\begin{align}
  \label{metksch}
  ds^2=\Bigl(1-\frac{2M}{a}\Bigr)dr^2-\Bigl(1-\frac{2M}{a}\Bigr)^{-1} da^2 -a^2 d\Omega^2 \,,
\end{align}
being valid for $a<2M$. Of course, $r$ is the \emph{temporal} coordinate, and $a$ turns out to be the radius of the 2-spheres. Once the interior metric is approximately established, we can enlarge the manifold by considering the region outside the event horizon. The result is the well known Schwarzschild metric in standard $(t,r,\theta,\phi)$ coordinates
\begin{align}
  \label{metkschfin}
  ds^2=\Bigl(1-\frac{2M}{r}\Bigr)dt^2-\Bigl(1-\frac{2M}{r}\Bigr)^{-1} dr^2 - r^2 d\Omega^2 \,.
\end{align}
Of course, neither (\ref{metksch}) nor (\ref{metkschfin}) are defined at the horizon itself; the KS metric (\ref{metksch}) describes the interior region only. In turn, (\ref{metkschfin}) is also unable to represent the horizon itself.

It is even more important to establish the correct structure of the vierbein field, which describes the actual dynamical field of $f(T)$ gravity. In order to do so, we need to perform the KS coordinate transformation on the proper frame (\ref{parasixs2}) taking into account the transformation law $e^b_{\mu'}=\partial x^{\mu}/\partial x^{\mu'}\,e^b_{\mu}$. Starting from (\ref{parasixs2}) and using the first of the equations (\ref{kschange4}) we obtain
\begin{align}
  \label{parasixs3}
  e^{(a)} = &-da \bigg/\sqrt{\frac{2M}{a}-1}\,,\\
  e^{(r)} = &\sqrt{\frac{2M}{a}-1}\,\cos\theta \,dr+a \sin^2\theta \,d\phi\,, \nonumber \\
  e^{(\theta)} = &\sqrt{\frac{2M}{a}-1}\,\sin\phi\sin\theta dr + a\sin\phi d\Omega_- \,, \nonumber \\
  e^{(\phi)} = &\sqrt{\frac{2M}{a}-1}\,\cos\phi\sin\theta dr - a\cos\phi d\Omega_+ \,,\nonumber
\end{align}
where the coordinates are $(a,r,\theta,\phi)$. This tetrad gives rise to the metric (\ref{metksch}), and represents -- approximately -- a parallelization of the interior black hole spacetime near the horizon. This frame field leads to a Weitzenb\"{o}ck scalar given by $T=4a^{-2}$, which is perfectly regular at the horizon $T_{\rm hor}=1/M^2$, even though the frame is not defined there.

In the same fashion as done before with the metric, we proceed to enlarge the tetrad structure in order to obtain a description of the asymptotically flat region lying far from the black hole horizon. However, this procedure is non trivial because we have to match tetrads $\mathcal{C}^{1}$ (at least once differentiable), i.e., it is necessary to $\mathcal{C}^{1}$ match four 1-form fields on a given hypersurface. This requirement is necessary because one has to guarantee a well posed dynamical evolution of system (\ref{ecuaciones}), which is a second order system of PDEs; the initial data will be constituted by the tetrad field and its first derivatives. This automatically will assure the continuity of the Weitzenb\"{o}ck scalar, which contains first derivatives of the tetrad field.

Guided by the metric (\ref{metkschfin}), we found
\begin{align}
  e^{(t)} &=\sqrt{1-\frac{2M}{r}}\,dt\,,\quad
  e^{(r)} = \frac{\cos\theta\,dr}{\sqrt{1-\frac{2M}{r}}}+r \sin^2\theta \,d\phi\,, \nonumber \\
  e^{(\theta)} &= \sin\phi \Bigl[\sin\theta dr/\sqrt{1-\frac{2M}{r}}+r d\Omega_- \Bigr]\,, \nonumber \\
  e^{(\phi)} &= \cos\phi \Bigl[\sin\theta dr/\sqrt{1-\frac{2M}{r}}-r d\Omega_+ \Bigr]\,.
  \label{parasixs4}
\end{align}
This frame leads to a scalar given by $T=4r^{-2}$ which, again, takes the value $T_{hor}=1/M^2$ at the horizon. In this way the scalars associated to both frames (\ref{parasixs3}) and (\ref{parasixs4}) have the structure $T=4/(radial\,\,coordinate)^2$; they lead to a $\mathcal{C}^{1}$ scalar all over the spacetime, even though at $r=2M$ (from the outside), or $a=2M$ (from the inside), the frames are not defined, just as the metric is undefined there. By changing to Eddington-Finkelstein or Kruskal coordinates in (\ref{parasixs3}) and (\ref{parasixs4}), the $\mathcal{C}^{1}$ character of the frame field can explicitly be shown.

The asymptotic flatness of the black hole reveals itself from (\ref{parasixs4}) taking the limit $r\rightarrow\infty$. In that limit the frame (\ref{parasixs4}) goes as
 \begin{align}
  e^{(t)}_{0} &\approx dt\,,\quad
  e^{(r)}_{0} \approx \cos\theta\,dr+r \sin^2\theta \,d\phi\,, 
  \\
  e^{(\theta)}_{0} &\approx \sin\phi [\,\sin\theta dr\,+r d\Omega_- ]\,,
  \nonumber
  \\
  e^{(\phi)}_{0} &\approx \cos\phi [\,\sin\theta dr\,-r d\Omega_+ ]\,,
  \nonumber
\end{align}
which leads to the line element $ds^2=dt^2-dr^2-r^2d\Omega^2$, characterizing Minkowski spacetime in spherical coordinates. More details about the nature of the proper frames in $f(T)$ gravity can be consulted in Refs.~\cite{grupo,NosUn}.

\section{Conclusions}
\label{conclusions}

The vacuum regular black hole studied here contains an interior region which is characterized asymptotically by a four dimensional extension of the non-rotating BTZ black hole. The BTZ horizon plays the role of future null infinity, in the small mass limit $\tilde{M} \ll 1$ where we recall that the BTZ mass $\tilde{M}$ is dimensionless. Moreover, the cosmological constant associated to the BTZ solution is $\Lambda=-\ell^{-2}$, i.e.,
\begin{align*}
  \Lambda=-\frac{\tilde{M}}{a_{c}^2(1+\tilde{M}/2)}=-\frac{\tilde{M}|\lambda|}{4(1+\tilde{M}/2)}\,,
\end{align*}
see expression (\ref{idenfin}). Consequently, the cosmological constant is also very small provided $\tilde{M}\ll 1$. The outer region, in turn, is indistinguishable from the Schwarzschild spacetime provided the Schwarzschild mass satisfies $M^2 \gg  1/|\lambda|$, or $r^2_{s} \gg a^2_{c}$. A freely falling observer trespassing the Schwarzschild horizon will asymptotically experience a constant curvature spacetime whose invariants are given by (\ref{escalares}). This asymptotic inner space is what we called a \emph{BTZ gem}, a sort of storage region which gathers the information carried by every form of matter/energy. The spacetime is then geodesically complete and free from the Schwarzschild singularity.

The thermodynamics when $M^2 \gg  |\lambda|^{-1}$ (early stage) is then the same as in the Schwarzschild case, even though corrections of order $|\lambda|^{-1}$ to the black hole entropy will certainly appear; these are of no concern for astrophysical and super-massive black holes and hence were not considered in the present work. However, the black hole temperature and entropy will differ radically from the Schwarzschild case when the mass approaches $M^2\approx|\lambda|^{-1}$, i.e., after a long evaporation time. The characterization of the final stages of the black hole evolution will involve a relation between the initial black hole mass $M$, the BI constant $\lambda$, and the BTZ mass $\tilde{M}$. These delicate matters are the subject of ongoing research.

\subsection*{Acknowledgments}
FF is member of Carrera del Investigador Científico (CONICET), and he is supported by CONICET and Instituto Balseiro. He is most thankful to J. Zanelli and CECS (Valdivia, Chile), where a stimulating discussion concerning this work took place.

The authors acknowledge support by The Royal Society, International Exchanges 2017, grant number \mbox{IEC\textbackslash R2\textbackslash 170013}.

\end{document}